**Electric energy by direct conversion from gravitational energy: a gift from superconductivity.**

**Osvaldo F Schilling**

Departamento de Física, Universidade Federal de Santa Catarina, Campus, Trindade, 88040-900, Florianópolis, SC. Brazil.

Email: osvaldof@mbox1.ufsc.br

**We theoretically demonstrate that electromagnetic energy can be obtained by direct, lossless, conversion from gravitational and kinetic energies. For this purpose we discuss the properties of an electromechanical system which consists of a superconducting coil submitted to a constant external force and to magnet fields. The coil oscillates and has induced in it a rectified electrical current whose magnitude may reach hundreds of Ampere. There is no need for an external electrical power source for the system to start out and it can be kept working continuously if linked to large capacitors. We extensively discuss the issue of energy dissipation in superconductors and show that the losses for such a system can be made extremely small for certain operational conditions, so that by reaching and keeping resonance the**

**system main application should be in magnetic energy storage and transmission.**

This paper is an extension of our previous work "**cond-mat/0308136**". The theory is discussed again here for the sake of clarity. A considerably more important application of the electromechanical oscillator is included in this version, alongside an extensive discussion about losses in type-II superconductors and on how to avoid them.

The demand for better electrical energy storage technology and cleaner power supplies has increased in recent years. Superconducting magnetic energy storage (SMES) technologies fulfil such requirements[1]. SMES usually involves the establishment of a permanent current in a solenoid which will keep the corresponding magnetic energy for future use. In this paper we discuss a somewhat different design for a SMES machine. Its operation principle is based upon the possibility of cyclic and conservative energy conversion between the lossless electronic motion inside a superconducting coil and the lossless mechanical oscillation associated with levitation of the coil in the presence of a system of magnets. Levitation itself is not obtained by the usual effect of repulsion against a dipolar field, like in other applications[2], but comes straight from the Lorentz forces acting upon the currents in the coil. Electrical currents are obtained from mechanical motion

33and vice-versa. The motion is linear, rather than rotatory, and no high-speed motion is necessarily involved. To achieve such electromechanical motion conversion the superconducting coil is submitted simultaneously to static uniform magnetic fields and to an external force. The imposition of an external static force is fundamental to the method in order to generate mechanical potential energy. For such conditions the coil will oscillate mechanically around an equilibrium position, and will have induced in it a rectified supercurrent containing an alternating component of same frequency. We will discuss the conditions necessary for such system to conserve its initial energy. Hundreds or even thousands of Ampere of rectified current may be obtained and the corresponding energy is kept stored in the magnetic field for future use.

Let´s consider the model experimental setup described in Fig. 1. A rectangular superconducting coil of mass $m$ is submitted to magnetic fields $B$ from a system of magnets. It is important that the entire coil be submitted to such fields, as discussed later. The coil is pulled away from the magnets by an external force $F$, which is assumed independent of time. According to Faraday's Induction Law the motion of the coil in the presence of $B$ will induce in it a transport supercurrent $i$. Such current will generate a magnetic force with strength $F_{m1}= iaB_1$ in the upper side ( of length $a$) of the coil in opposition to *F,* and another force of strength $F_{m2}= iaB_2$ in the lower side. Here $B_1$ and $B_2$ are the magnetic fields produced by the magnets at the position of the sides of length *a*. The magnetic forces must not cancel each other otherwise no effect is obtained. We will assume $B_1 > B_2 > B_{c1}$, where $B_{c1}$ is the lower critical field of the superconducting material. The coil will move with speed *v* described by Newton´s Law:

$$m\, dv/dt = F - ia(B_1 - B_2) \qquad (1)$$

We introduce the parameter $B_0 \equiv B_1 - B_2$ to simplify the notation. The displacement of the coil gives rise to an induced electromotive force $\varepsilon$, given by Faraday's Induction Law[3]:

$$\varepsilon = -d\Phi_m/dt - L\, di/dt \qquad (2)$$

Here $\Phi_m$ is the magnetic flux from the magnets that penetrates the rectangular area bound by the coil, and *L* is the self-inductance of the coil. The electromotive force may be considered zero for all practical purposes, since the wire is superconducting, and this results in the conservation of the total magnetic flux ($\Phi_m + L\, i$) inside the area bound by the coil. In (2), $d\Phi_m/dt = -B_0 av$. It is important to stress the absence of dissipative terms in (2), a hypothesis that will be extensively discussed later. From (2) one obtains a relation between *v* and $di/dt$. Taking the time derivative of (1) and eliminating $di/dt$ from (2) one obtains:

$$m\, d^2v/dt^2 = -(B_0^2 a^2/L)\, v \qquad (3)$$

This means that the coil shall perform an *ideal* oscillating motion under the action of the external and magnetic forces. Assuming zero initial speed and an initial acceleration equal to $F/m$ eq. (3) can be solved:

$$v(t) = F/(m\Omega)\, \sin(\Omega t) \qquad (4)$$

The natural frequency of the oscillations is $\Omega = B_0 a/(mL)^{1/2}$. The amplitude of the oscillating motion is $x_0 = F/(m\Omega^2)$, which may be quite small since it is inversely proportional to $\Omega^2$. It is possible then to combine (1)–(2) to obtain an equation for the current $i(t)$:

$$(B_0 a/\Omega^2)\, d^2 i/dt^2 = F - B_0 a i \qquad (5)$$

whose solution is

$$i(t) = (F/(B_0 a))\, (1 - \cos(\Omega t)) \qquad (6)$$

for $i(0) = di/dt(0) = 0$. From eq. (6) we conclude that the supercurrent induced in the coil is already rectified. It never changes sign, and it looks like the result of submitting an alternating current of frequency $\Omega/2$ to a diode rectifying bridge. Typical figures for the speed amplitude ($v_o$), oscillation amplitude, and rectified current amplitude ($i_o = F/(B_0 a)$) may be obtained. Let's take $L = 10^{-7}$ H, $B_0 = 0.3$ T, $m = 1$ kg, $F = 10$ N, $a = 0.1$ m. For these parameters, $\Omega = 95$ rad/s, $i_o = 333$ A, $v_o = 11$ cm/s, and $x_o = 1.1$ mm. That is, a large rectified current may be obtained with low speed and frequency of oscillation, and very small displacements of the coil.

The set of eq. (1)-(6) was deduced under the assumption of the absence of energy losses. Let's discuss the conditions for such assumption to apply. First of all, it is important to stress that the predicted mechanical motion can be made perfectly frictionless with this design, since in the vertical position the motion is independent of any physical contact between the levitating coil and the magnet. If friction losses are negligible, the next issue that immediately arises is that of the possible losses related to the alternating current in the wire and to inductive coupling between the coil and the magnets. Inductive coupling between the coil and a conducting magnet like Nd-Fe-B will produce resistive eddy currents in the magnet[2]. This source of dissipation might be eliminated by using an insulating magnet like ferrite, at the possible cost of having to work at lower fields. Assuming that this source of energy dissipation may be circumvented let's discuss how to avoid losses associated with the transport of ac currents by the superconductor wire[4-6]. Such losses may be classified in resistive losses and hysteretic losses. Resistive losses might arise due to partial current transportation by normal electrons. The influence of normal electrons may be avoided by working at low frequencies[4,5]. We note that the number of free parameters of the model allows the frequency $\Omega$ to be set, e.g., in the 10 ~ 1000 rad/s range, so that the influence of eddy current losses due to normal carriers expected already in the upper MHz range[5] may be entirely neglected. Alternating currents give rise to a resistive response from the superconducting electrons also[6]. However, well below $T_c$ such resistivity is smaller than the normal state resistivity by a factor $\hbar\Omega/\Delta$, where $\Delta$ is the energy gap[6]. Such factor is of the order of $10^{-9}$ for low frequencies, so that the resistive behavior of superconducting electrons may safely be neglected. Hysteresis losses occur whenever the flux line (FL) lattice inside a type-II superconductor is cyclically rebuilt by an oscillating magnetic field[4]. Such fields ( ripple fields) are created by the alternating part of the current. The work of Campbell, Lowell, and others[7-9] has shown that when the FL are displaced from their pinning sites by a distance $d_o$ as small as the coherence length ( 2 ~ 6 nm) such displacements are elastic and reversible, and *no* hysteresis losses occur. From the technical point of view it is very important to know which magnitude of ac current will produce such threshold elastic displacements of the FL in a given material. Campbell's work[7] has shown that the threshold ripple field is related to the spacing $D$ between microstructural features that provide flux-pinning[10]. Let $b$ represent the ripple-field amplitude, and $B$ the static magnetic field. Campbell[7] has shown that there will be no dissipation if (approximately) $b/B \leq d_o/D,$ with $d_o$ given above. This explains why the entire coil should be submitted to the static magnetic fields. To avoid hysteresis losses due to the ripple field produced by the ac current the value of $B$ must be large compared to $b$, otherwise the $b/B \leq d_o/D$ criterion will not be obeyed. In the case of a Pb-Bi eutectic sample[7], pinning was provided by Bi particles of 5 µm size. Campbell's measurements displayed *no* hysteresis losses for $b = 10^{-4}$ T, $B = 1.1$ T, and a value of $D$ of order 20 µm, which allowed the determination of $d_o = 2$ nm. This suggests that if stronger pinning materials like Nb-Ti wires are adopted much greater internal ripple fields, on the order of 0.1 T, for instance, might be sustained without losses. Just to give a numerical example, if a cold-drawn Nb-Ti wire 0.17 mm thick[11] is used to make the coil and the alternating current amplitude is 100 A the ripple field produced by the wire at its outer surface will be of the order of 0.1 T. Since for such material[11] the distance $D$ between pinning Ti particles is about 20 nm, one might be able to impose a maximum ripple field $b = d_o B/D = 2/20 = 0.1$ T for a static field $B=1$ T and $d_o = 2$ nm. That is, an alternating current of 100 A ( corresponding in this example to a very high current density of $5 \times 10^9$ A/m$^2$) might be sustained without hysteresis losses provided the coil material is properly chosen.

An additional source of hysteresis might come not from the effect of the ripple fields upon the FL lattice, but by the motion of the coil in the presence of static fields whose magnitude varies in space. That is the situation shown in Fig.1, in which two magnets (should) provide homogeneous magnetic fields of different strengths $B_1$ and $B_2$. The way of minimizing such losses is by restricting the coil oscillation amplitude to very low values, so that the extension of coil in which the flux density actually changes during cycles is as small as possible. In the example given above the amplitude of oscillation was only 1 mm, which is probably of comparable magnitude to the finest scale of homogeneity provided by the best magnets. One might mention also flux-creep as another possible source of dissipation, but the use of a temperature of operation well below $T_c$ , and a strong-pinning material should discard the possibility of creep.

Having discussed the several possible sources of dissipation, it is possible to conclude that losses may be severely restricted, if not fully eliminated, by a proper choice of magnet design and magnetic material together with a proper design of the coil and operation conditions. Nevertheless, it seems important also to quantitatively estimate the effects of external loads upon the behavior of the system. Let's consider first the effect of adding a resistance $R$ to the circuit. In this case equation (2) becomes:

$$0 = d\Phi_m/dt + L\, di/dt + R\, i \qquad (7)$$

Following the same procedure as before it is possible to calculate both current and speed as a function of time by solving eqs. (1) and (7). The speed ceases to be purely oscillating, displaying an additional component that is initially a linear function of time and tends to a saturated maximum value at long times. The current also has its oscillations damped and tends to the constant value $F/(B_0 a)$ at long times. Of course this is a consequence of energy losses in the resistor, which will make the coil progressively escape from inside the magnets region under the effect of $F$. It is very important, however, that the coil reaches resonance ( even if for just a brief period of time) since this will allow the induced current to reach its maximum value ( together with the stored magnetic energy). The oscillating component of the speed is dominant provided $R/L < 2\Omega$. The current produced will still be rectified, but its frequency becomes $\omega_0 = (\Omega^2 - R^2/(4L^2))^{1/2}$. The frequency must be a real number and this imposes a limit on the maximum stray resistance that the system may sustain. For the numerical example given before $R$ should be smaller than $2 \times 10^{-5}$ ohm, but preferably much smaller than this. This shows how important is the proper control of dissipation for the operation of this system. A similar effect is obtained by adding a capacitor $C$ to the oscillator circuit. The oscillator will behave like a "charge pump", transferring charge and energy to the capacitor. Equation (2) becomes:

$$0 = d\Phi_m/dt + L\, di/dt + q/C \tag{8}$$

The current produced is rectified and *undamped*, but its frequency becomes $\omega_{0c} = (\Omega^2 + 1/(LC))^{1/2}$. The capacitor will continuously be loaded, so that the time-averaged potential difference $V$ between its plates increases at a rate $dV/dt = (F/(B_0 aC))(\Omega/\omega_{0c})^2$. Again the speed of the coil displays a component $v_c$ that varies linearly with time. Such speed $v_c$ is *inversely* proportional to $C$ for very large values of $C$, so that if a very large capacitance bench is linked to the oscillator energy transfer can take place continuously with little effect on the oscillator motion.

    In resume, the main effects of the energy losses in the system is the progressive escape of the coil from the magnets region, and alterations or even the disappearance of the oscillations. These detrimental effects may be avoided by the proper control of losses in the design. An important operational detail is that as long as the coil oscillates inside the magnets region there will be a maximum current flowing, and the stored energy will still reach its maximum. Therefore, it seems recommendable that the system includes a device that will reverse the direction of the force $F$ at fixed time intervals. This will help to avoid the escape of the coil from the magnets area in case of losses, allowing the system to operate in continuous mode.

In conclusion, the present work has described the principles of design of a "stand alone" SMES machine, which will generate electromagnetic energy by conversion from kinetic and potential gravitational energies, with no need for an external power supply for it to start out. The system might be linked to a large capacitance bench (thousands of Farad or even more) in order that its energy can be continuously transferred for other applications. We have extensively discussed the conditions required to avoid dissipative effects due to normal resistivity, hysteresis, and flux-creep, so that the system may operate in a continuous mode. The actual implementation of a system with these characteristics seems perfectly possible in view of the available materials and magnet fabrication technologies.

The author wishes to thank Prof. Said Salem Sugui Jr. and Prof. Mauro M. Doria for helpful discussions.

**References.**

[1] Ries G and Neumueller H-W 2001 *Physica C* **357-360** 1306.

[2] Brandt E H 1988 *Appl. Phys. Lett.* **53** 1554.

[3] Chandrasekhar B S, in Parks RD ( editor) 1969 *Superconductivity*(New York: M. Dekker), Volume 1, chapter 1.

[4] Wilson MN 1989 *Superconducting Magnets* (Oxford: Oxford University Press).

[5] McLean W L 1962 *Proc. Phys. Soc. (London)* **79** 572.

[6] Tinkham M 1980 *Introduction to Superconductivity* ( Malabar, Fl. : Krieger) pp. 68-71.

[7] Campbell A M 1971 *J. Phys. C* **4** 3186.

[8] Lowell J 1972 *J. Phys. F* **2** 547 and 559 .

[9] Seow WS *et al*. 1995 *Physica C* **241** 71.

[10] We have adopted this interpretation for $D$ since it is convenient for obtaining numerical estimates for the threshold ripple field $b$ from the knowledge of the wire microstructure. For a more detailed analysis of this point see page 3194 of ref. 7.

[11] Meingast C Lee PJ and Larbalestier D C 1989 *J. Appl. Phys.* **66** 5962.

Figure caption.

Figure 1: A rectangular coil is submitted simultaneously to an external force $F$ and magnetic fields $B_1$ and $B_2$ perpendicular to it. As shown in the text the predicted motion is oscillatory, with an alternating current of same frequency being induced in the coil.

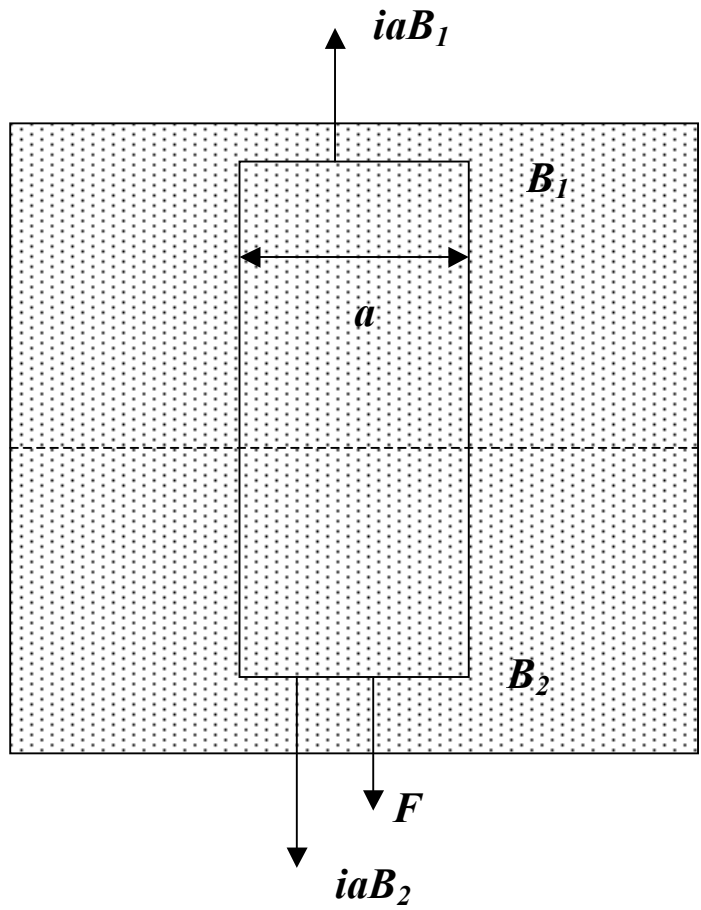

Figure 1
Schilling